\documentclass[]{aa}  
\usepackage{subfigure}
\usepackage{graphicx}
\usepackage{txfonts}
\usepackage{natbib,twoopt}
\usepackage[breaklinks=true]{hyperref} 
\bibpunct{(}{)}{;}{a}{}{,}             
\makeatletter
  \newcommandtwoopt{\citeads}[3][][]{\href{http://adsabs.harvard.edu/abs/#3}%
    {\def\hyper@linkstart##1##2{}%
     \let\hyper@linkend\@empty\citealp[#1][#2]{#3}}}
  \newcommandtwoopt{\citepads}[3][][]{\href{http://adsabs.harvard.edu/abs/#3}%
    {\def\hyper@linkstart##1##2{}%
     \let\hyper@linkend\@empty\citep[#1][#2]{#3}}}
  \newcommandtwoopt{\citetads}[3][][]{\href{http://adsabs.harvard.edu/abs/#3}%
    {\def\hyper@linkstart##1##2{}%
     \let\hyper@linkend\@empty\citet[#1][#2]{#3}}}
  \newcommandtwoopt{\citeyearads}[3][][]%
    {\href{http://adsabs.harvard.edu/abs/#3}
    {\def\hyper@linkstart##1##2{}%
     \let\hyper@linkend\@empty\citeyear[#1][#2]{#3}}}
\makeatother

\begin{document}

   \title{Detecting Galaxy Groups and AGNs populating the local Universe \\in the eROSITA era}

   \author{I. Marini\thanks{ilaria.marini@eso.org}
          \inst{1}
          \and
          P. Popesso\inst{1}$^,$\inst{2}
          \and
          G. Lamer\inst{3}
          \and
          K. Dolag\inst{4}$^,$\inst{5}$^,$\inst{2}
          \and 
          V. Biffi\inst{6}$^,$\inst{7}
          \and 
          S. Vladutescu-Zopp\inst{4}
          \and
          A. Dev\inst{10}$^,$\inst{11}
          \and
          V. Toptun\inst{1}
          \and
          E. Bulbul\inst{8}
          \and
          J. Comparat\inst{8}
          \and
          N. Malavasi\inst{8}
          \and
          A. Merloni\inst{8}
          \and
          T. Mroczkowski\inst{1}
          \and
          G. Ponti\inst{9}$^,$\inst{8}
          \and
          R. Seppi\inst{12}   
          \and
          S. Shreeram\inst{8} 
          \and
          Y. Zhang\inst{8}
          }

   \institute{European Southern Observatory, Karl-Schwarzschildstr. 2, 85748, Garching bei M\"unchen, Germany\\
            \email{ilaria.marini@eso.org}
         \and
            Excellence Cluster ORIGINS, Boltzmannstr. 2, D-85748 Garching bei M\"unchen, Germany
        \and
            Leibniz-Institut für Astrophysik Potsdam (AIP), An der Sternwarte 16, 14482 Potsdam, Germany
        \and
             Universitäts-Sternwarte, Fakultät für Physik, Ludwig-Maximilians-Universität München, Scheinerstr.1, 81679 München, Germany
        \and 
            Max-Planck-Institut für Astrophysik, Karl-Schwarzschildstr. 1, 85741 Garching bei M\"unchen, Germany
        \and
             INAF – Osservatorio Astronomico di Trieste, Via Tiepolo 11, 34143 Trieste, Italy
        \and 
            IFPU – Institute for Fundamental Physics of the Universe, Via Beirut 2, I-34014 Trieste, Italy
        \and
            Max-Planck-Institut für Extraterrestrische Physik (MPE), Giessenbachstr. 1, D-85748 Garching bei München, Germany
        \and
            INAF– Osservatorio Astronomico di Brera, Via E. Bianchi 46, 23807 Merate (LC), Italy
        \and
            International Centre for Radio Astronomy Research, University of Western Australia, M468, 35 Stirling Highway, Perth, WA 6009, Australia
        \and
            ARC Centre of Excellence for All Sky Astrophysics in 3 Dimensions (ASTRO 3D), Australia
        \and
            Department of Astronomy, University of Geneva, Ch. d’Ecogia 16, CH-1290 Versoix, Switzerland
             }

   \date{Received  ; accepted  }
 
  \abstract
   {The extended ROentgen Survey with an Imaging Telescope Array (eROSITA) will deliver an unprecedented volume of X-ray survey observations, $20-30$ times more sensitive than ROSAT in the soft band ($0.5-2.0$ keV) and for the first time imaging in the hard band ($2-10$ keV). The final observed catalogue of sources will include galaxy clusters and groups along with obscured and unobscured Active Galactic Nuclei (AGNs). This calls for a powerful theoretical effort to control the systematics and biases that may affect the data analysis. }
   {We investigate the detection technique and selection effects in the galaxy group and AGN populations of a mock eROSITA survey at the depth of a four-year eROSITA survey (eRASS:4).}
   {We create a $30\times 30$ deg$^{2}$ mock observation based on the cosmological hydrodynamical simulation {\it Magneticum Pathfinder} from $z=0$ up to redshift $z=0.2$ at the depth of eRASS:4 (average exposure $\sim 600$ s). We combine a physical background extracted from the real eFEDS background analysis with realistic simulations of X-ray emission for the hot gas, AGNs and X-ray binaries. We apply a detection procedure equivalent to the reduction done on eRASS data and evaluate the completeness and contamination to reconstruct the luminosity functions of the extended and point sources in the catalogue. }
   {We assess the completeness of extended detections as a function of the input X-ray flux $S_{500}$ and halo mass $M_{500}$ at the depth of eRASS4. Notably, we achieve full recovery of the brightest (most massive) galaxy clusters and AGNs. However, a significant fraction of galaxy groups ($M_{200}<10^{14} M_{\odot}$) remains undetected. Examining the gas properties between the detected and undetected galaxy groups at fixed halo mass, we observe that the detected population exhibits, on average, higher X-ray brightness compared to the undetected counterparts. Moreover, we find that X-ray luminosity primarily correlates with the hot gas fraction, rather than temperature or metallicity. Our simulation suggests the presence of a systematic selection effect in current surveys, resulting in X-ray survey catalogues predominantly composed of the lowest-entropy, gas-richest, and highest surface brightness halos on galaxy group scales.} 
   {}


   \maketitle
%

\section{Introduction}
Despite their abundance in the Universe \citep{tinker_toward_2008}, galaxy groups ($M_{200}$\footnote{We can define $M_{\Delta}$ as the mass encompassed by a mean overdensity equal to $\Delta$ times the critical density of the universe $\rho_c(z)$.} $\sim 10^{12.5}-10^{14} M_{\odot}$) have only recently been accessible in X-rays for systematic studies of their hot gas properties. Their shallow potential wells and faint X-ray emissions cause several challenges in detecting these systems, even though a large fraction of the baryonic mass resides here \citep{eckert_feedback_2021}. In contrast to the hot galaxy clusters, groups are systems where the baryon physics -- e.g., cooling, galactic winds and Active Galactic Nuclei (AGN) feedback -- begins to dominate \citep{eckert_feedback_2021}. They are not simply scaled-down versions of massive clusters and recently these claims have been sustained by many scaling relations not holding up at all mass scales \citep[see discussion in][]{lovisari_scaling_2021, oppenheimer_simulating_2021}. The IntraGroup Medium (IGrM) primarily emits due to bremsstrahlung interactions, however, cooling through line emission is significant and it can lead to the presence of multi-phase gas in the same environment \citep[e.g.,][]{zuhone_effects_2023}. For this reason, X-ray observations should trace fairly well the IGrM and provide a meaningful understanding of the thermodynamic properties of the hot gas \citep[see][for a review]{voit_global_2017, lovisari_scaling_2021}. Their low surface brightness may be misclassified as pointed emission due to redshift dimming can only be resolved through multi-wavelength follow-up observations \citep{salvato_erosita_2022, bulbul_erosita_2022}. 
\par
In this context, the extended ROentgen Survey with an Imaging Telescope Array (eROSITA) on board SRG will deliver the first census of the entire X-ray sky with high sensitivity in the soft X-ray band and a scanning observing strategy \citep{merloni_erosita_2012, sunyaev_srg_2021}. The space telescope will overcome previous limits on the statistics, depth and spatial resolution allowing us to increase the volume of our X-ray surveys by at least one order of magnitude in the groups/clusters regime and reach a total of 3 million AGNs caught at different evolutionary phases and clustering up to $z=2$ \citep{merloni_erosita_2012}. Its scientific goals range from understanding purely astrophysical phenomena (e.g., characterising the thermal structure and chemical enrichment of halos as a function of redshift) to providing constraints on the cosmological parameters \citep{ghirardini_srgerosita_2024, artis_srg-erosita_2024}, tasks that have already shown promising results with the data releases of the German sky in the eROSITA Final Equatorial Depth Survey \citep[eFEDS][]{brunner_erosita_2022} and eROSITA All-Sky Survey-1 \citep[eRASS1][]{merloni_srgerosita_2024}.  eROSITA will deliver the first large statistical determination of cross-correlations between AGNs-clusters and AGNs-groups \citep{cappelluti_soft_2007}, thus it will be key to understanding the interplay between the AGN feedback and the hot gas permeating the group environment \citep[e.g.,][]{bahar_erosita_2022, bahar_srgerosita_2024}.
\par
Such an attractive challenge calls for a joint effort on both the theoretical and observational sides to benchmark the systematics and limitations of our observations. The first step will be to define a clear selection function for a survey like the one delivered by eROSITA (i.e., the completeness and selection effects) and secondly understand to what extent background and foreground sources might contaminate the data. \cite{clerc_synthetic_2018, comparat_full-sky_2020, liu_establishing_2022, seppi_detecting_2022, clerc_srgerosita_2024} addressed the issue with a full-sky dark matter (DM) only lightcone on which they painted galaxy properties and X-ray emission based on the most recent observational data. This allowed them to infer the expected X-ray selection function at different exposure times. Nevertheless, these studies lack a self-consistent modelling of the baryon physics in the galaxy population with the environment \citep{comparat_full-sky_2020}. From the hydrodynamical simulations part, there have been several efforts to provide predictions for the hydrostatic mass bias \citep{scheck_hydrostatic_2023}, projection effects \citep{zuhone_effects_2023} and the gas properties observed in nearby clusters \citep{biffi_erosita_2022, churazov_prospects_2023}. However, we still lack a comprehensive study on the halo population and the systematics driving many of these objects to be undetected \citep{popesso_x-ray_2023}.
Furthermore, many of these cosmological simulations rely on different feedback prescriptions \citep{hirschmann_cosmological_2014, vogelsberger_introducing_2014, steinborn_refined_2015, schaye_eagle_2015, dave_simba_2019, zinger_ejective_2020} which are expected to be most effective at these scales \citep{fabian_observational_2012, gaspari_linking_2020, eckert_feedback_2021}: investing our efforts in comparing the observational data with the simulations will provide the necessary tools to constrain our models. It is interesting to notice that some recent simulations, including the IllustrisTNG project \citep{pillepich_simulating_2018, nelson_first_2018}, SIMBA \citep{dave_simba_2019} and FLAMINGO \citep{schaye_flamingo_2023}, are explicitly calibrated to reproduce several observed scaling relations down to the group-scale halos. This calibration may lead to biased results since our goal is to test these same gas halo properties. To overcome this issue, we create mock observations with the state-of-the-art \textit{Magneticum Pathfinder} \citep{dolag_sz_2016} which is not explicitly calibrated against gas properties. To simulate the telescope's scanning strategy and characteristics we make use of SIXTE \citep{dauser_sixte_2019} the official end-to-end simulator for eROSITA.
\par
This is the first of a series of papers testing this framework with a particular focus on the galaxy group regime in the local Universe.  We construct eROSITA mock observations of a patch of the sky ($30 \times 30$ deg$^{2}$) up to redshift $z=0.2$ using the scanning strategy designed for eRASS:4. We reconstruct the selection functions and pave the way for a series of works focused on the comparison between simulations and observations. Future papers will involve the analysis at higher redshifts for which we create eight lightcones down to $z=1.1$ but with a smaller sky area (i.e., $5\times 5$ deg$^2$).
\par
The paper is structured as follows. In Sect.~\ref{sec:simulations} we present the set of simulations and the lightcone designed for this experiment. Sect.~\ref{sec:X-ray_catalogue} describes the spectral modelling of the X-ray emission along with the techniques applied to obtain a mock catalogue. Sect.~\ref{sec:eSASS_catalogue} illustrates the matching technique to recover the detected and undetected underlying halo population in the simulations, this will pose the necessary base to reconstruct the selection function. The detected population is used to reconstruct the luminosity function to compare it with the eFEDS luminosity function in Sect.~\ref{sec:luminosity_function}. Sect.~\ref{sec:undetected} elaborates more on the selection effects in place which lead (at fixed halo mass) the detection/non-detection of galaxy groups. Finally, in Sect.~\ref{sec:conclusions}, we will conclude our study and provide a summary.

\section{Simulations}
\label{sec:simulations}
The {\it Magneticum Pathfinder} simulation\footnote{\url{http://www.magneticum.org/index.html}} is a large set of state-of-art cosmological hydrodynamical simulations carried out with P-GADGET3, an updated version of the public GADGET-2 code \citep[][]{springel_cosmological_2005}. The most important changes comprehend a higher-order kernel function, time-dependent artificial viscosity and artificial conduction schemes \citep{dolag_turbulent_2005,beck_improved_2016}. 
\par
Several subgrid models describe the unresolved baryonic physics of the simulations, including radiative cooling \citep{wiersma_effect_2009}, a uniform time-dependent UV background \citep{haardt_modelling_2001}, star formation and stellar feedback \citep[i.e., galactic winds][]{springel_cosmological_2003}, chemical enrichment  (following explicitly H, He, C, N, O, Ne, Mg, Si, S, Ca, Fe) due to stellar evolution \citep{tornatore_chemical_2007}. Models for SuperMassive Black Hole (SMBH) growth and accretion and AGN feedback are included following the prescription in \cite{springel_cosmological_2005,di_matteo_energy_2005, fabjan_simulating_2010, hirschmann_cosmological_2014}. 
\par
Halos and substructures in the simulations are identified in post-processing firstly applying a Friend-of-Friend (FOF) algorithm and later SubFind \citep{springel_populating_2001, dolag_substructures_2009}. 
\par
Many studies have already shown the capabilities and quality of the predictions of \textit{Magneticum} when compared to the observational data. Without the intent to be comprehensive, we list here the comparison with galaxy properties \citep{teklu_connecting_2015, teklu_dynamical_2016, teklu_morphology-density_2017, remus_outer_2017, schulze_kinematics_2017, schulze_kinematics_2018, schulze_kinematics_2020}, galaxy clusters \citep{biffi_investigating_2013, dolag_distribution_2017, ragagnin_simulation_2022} and X-ray emission of galaxies and galaxy clusters \citep{biffi_investigating_2013, biffi_erosita_2022, veronica_erosita_2022, veronica_erosita_2024, seppi_offset_2023, scheck_hydrostatic_2023, vladutescu-zopp_decomposition_2023, bogdan_circumgalactic_2023, zuhone_effects_2023, churazov_prospects_2023, bahar_srgerosita_2024}. Therefore in this work, we use the run from \textit{Box2/hr} which follows the evolution of a large ($352\, h{^{-3}}$ cMpc$^{3}$) cosmological box with $2\times15843$ particles at the following resolutions: $m_\mathrm{DM} = 6.9\times10^8\, h^{-1}$~M$_{\odot}$ and $m_\mathrm{gas} = 1.4\times10^8\, h{^{-1}}$~M$_{\odot}$. 
The Plummer equivalent length for the DM particles corresponds to $\epsilon=3.75\, h{^{-1}}$ kpc, whereas gas, stars and black hole particles retain $\epsilon=3.75\,h^{-1}$~kpc, $2\, h{^{-1}}$~kpc and $2\,h{^{-1}}$~kpc at $z=0$, respectively.
\par
The simulation is run with the cosmology from WMAP7 measurements \citep{komatsu_hunting_2010} $\Omega_\mathrm{M}=0.272$, $\Omega_\mathrm{b}=0.0168$, $n_s=0.963$, $\sigma_8 = 0.809$ and $H_0=100 \, h$ km~s${^{-1}}$~Mpc${^{-1}}$ with $h=0.704$. 
\begin{figure}
    \centering
    \includegraphics[scale=0.41]{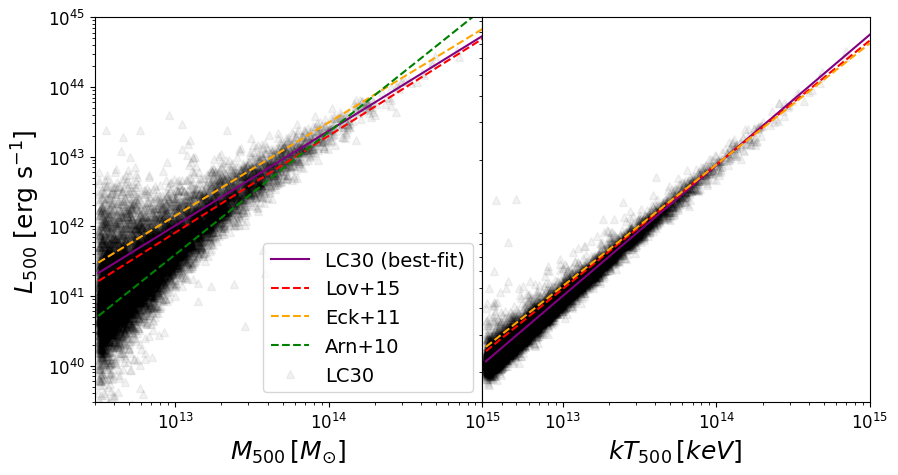}
    \caption{X-ray scaling relations recovered from the halo population in the lightcone: in the left panel $L_{500}-M_{500}$ and right panel $L_{500}-kT_{500}$. Notice that the luminosity $L_{500}$ is intended as the IGrM-only emission including Galactic absorption in the $0.1-2.4$ keV ban within $R_{500}$. For reference, also literature values are reported: Lov+15 \citep{lovisari_scaling_2015}, Eck+11 \citep{eckmiller_testing_2011}, Arn+10 \citep{arnaud_universal_2010}. The masses and mass-weighted temperature are calculated in the same radius $R_{500}$. Considering the small redshift range, we neglect any evolutionary changes in the scaling. }
    \label{fig:scaling_relations}
\end{figure}
\subsection{Setup of the lightcone}
The lightcone is designed by extracting random sub-cubes at five different simulation snapshots and combining them according to the preferred geometrical configuration from the most recent to the furthest back. Each sub-cube is extracted to provide the necessary continuity in the selected angular aperture and depth. This design poses two main limitations to our ability to create an arbitrary number of lightcones from the same simulation. On the one hand, the choice of depth in redshift $z$ and angular aperture $\theta$ of the lightcone are intertwined and limited by the volume of the cosmological box (i.e., a cube of size $L_\mathrm{Box}$, thus  $L^3_\mathrm{Box}$). This is a consequence of the geometrical distribution of the sub-cubes since the maximum angular aperture $\theta_\mathrm{max}$ allowed is
\begin{equation}
    \theta_\mathrm{max} = \frac{L_\mathrm{Box}}{w(z_\mathrm{max}, \Vec{x})}
\end{equation}
with $w(z, \Vec{x})$ being the angular distance at redshift $z$ in a universe with cosmological parameters $\Vec{x}$. On the other hand, there is only a finite number of independent realisations from one single box depending on both the specified geometrical configuration (i.e., the maximum sub-cube size $L_\mathrm{max}$) and the total volume of the box, namely, 
\begin{equation}
    N_\mathrm{max} = 3 \times \, \mathrm{int} \left( \frac{L_\mathrm{max}}{L_\mathrm{Box}}\right ).
\end{equation}
Therefore, we carefully choose to extract 1 lightcone of size 30 deg $\times$ 30 deg (LC30). In each sub-cube, the particles (and accordingly, the halos and galaxies) are redshifted/blueshifted according to the nominal distance from the observer located at $(0,0,0)$. For each halo, we compute the corresponding sky (angular) coordinate in the equatorial system -- centred on (RA, Dec)~=~(0,0) -- along with the true and the observed redshifts deduced by the peculiar velocity of the galaxies. Accounting for such calculation allows us to analyse the lightcone in redshift-space where the distribution of galaxies is distorted with the \textit{"finger-of-God"} effect \citep{zehavi_galaxy_2002}. This is not the first time lightcones were generated from the {\it Magneticum} output: \cite{dolag_sz_2016} investigated the thermal and kinetic Sunyaev-Zeldovich effect and the mean Compton Y parameter in comparison with the Planck, South Pole Telescope (SPT) and the Atacama Cosmology Telescope (ACT) data.

\section{Mock X-ray observation}
\label{sec:X-ray_catalogue}
In this section, we explain the steps for producing the eROSITA mock observation based on  \textit{Magneticum}. This is done through the following steps:
\begin{enumerate}
    \item The lightcones are processed through PHOX \citep{biffi_observing_2012, biffi_investigating_2013, biffi_agn_2018} to produce an ideal event list based on the physical properties of the gas, BH and stellar particles in the simulation (see Sect.~\ref{subsec:3.1}). 
    \item The event file is run by SIXTE \citep{dauser_sixte_2019} to extract a mock observation in scanning mode as eFEDS and eRASS:4 (see Sect.~\ref{subsec:3.2}). 
    \item The resulting mock observation, including all eROSITA instrumental effects and calibrations, are processed through the eROSITA Science Analysis Software System (eSASS) to extract extended and point source detections as done in \cite{merloni_srgerosita_2024, bulbul_srgerosita_2024}  (see Sect.~\ref{subsec:3.3}).
\end{enumerate}
A more in-depth description of each step is given in the following.
\subsection{Creating the event file with PHOX}
\label{subsec:3.1}
The eROSITA X-ray emission is obtained by modelling the hot gas, AGNs and X-ray binaries in the simulations.
Firstly, we generate a photon list using \texttt{PHOX} \citep{biffi_observing_2012, biffi_investigating_2013}. The code computes X-ray spectral emission for the aforementioned components based on the physical properties of the gas, BH and stellar particles in the simulation.
A fiducial collecting area $A_\textrm{fid}$ and exposure time $\tau_\textrm{fid}$ are initially assumed to sample a discrete and ideally large number of photons from the spectra in a Monte Carlo approach. In the second step, \texttt{PHOX} considers the lightcone's geometry and projects the emission along the line of sight while photon energies are corrected for the Doppler shift \citep[see Unit2 in][]{biffi_agn_2018}. This second unit follows the prescription derived from the lightcone geometry (e.g., the field of view, and position in the cosmological box) to be consistent throughout. Each component has a different spectral emission modelled with the XSPEC library \citep[v12;][]{arnaud_xspec_1996}.
\paragraph{Modelling the hot gas emission.} The hot gas emission is assumed to follow \texttt{vapec} \citep{smith_collisional_2001} with a single temperature model and in the presence of heavy elements, for which single abundances are explicitly tracked. Solar abundances are assumed \cite{anders_abundances_1989}. The parameters are derived from the chemical and thermal properties of the single gas particles in the simulations.  A simple foreground absorber is assumed with the \texttt{wabs} model \citep{morrison_interstellar_1983} with column density $N_{H}=10^{20}$ cm$^{-2}$. We refer to the description in \cite{biffi_observing_2012,biffi_investigating_2013} for further details.
\paragraph{Modelling the AGN emission.} The synthetic X-ray spectrum of AGNs is modelled in PHOX from the properties of BH particles in the simulation with an intrinsically absorbed power-law, having assigned an obscuring torus to every source. The absorber's column density is randomly drawn from the distribution described in \cite{buchner_x-ray_2014}. More details on the implementation can be found in \cite{biffi_agn_2018}. Since then, the absorption model has changed to the \texttt{tbabs} photoionization cross sections \citep{wilms_absorption_2000}, instead of the \texttt{wabs} cross sections \citep{morrison_interstellar_1983}. 
\paragraph{Modelling the XRB emission.} The emission of XRBs in PHOX, which is stellar in origin, is traced with the star particles in the simulation. The spectral shape is an absorbed (\texttt{ztbabs}) power-law (\texttt{zpowerlw}) with a photon index extracted according to the XRB type. The XRBs can be classified according to the predominant accretion process: wind-fed in the case of high-mass XRBs and Roche-lobe mass transfer in the case of low-mass XRBs. The HMXB has photon index $\Gamma=2$ \citep{mineo_x-ray_2012}, and the LMXB has $\Gamma=1.7$ \citep{lehmer_x-ray_2019}. The full description of the XRB emission modelling in PHOX is presented in~\cite{vladutescu-zopp_decomposition_2023}.
In Fig.~\ref{fig:scaling_relations}, we show two scaling relations obtained from the IGrM photons compared to several observational best-fit relations, namely \cite{lovisari_scaling_2015, eckmiller_testing_2011, arnaud_universal_2010}. We calculate these quantities within a spherical aperture with radius $R_{500}$ centred on the particle at the minimum of the gravitational potential. 

\subsection{Event files}
\label{subsec:3.2}
The synthetic photon lists are the input files to the Simulation of X-ray Telescopes (SIXTE) software package \citep[v2.7.2]{dauser_sixte_2019}, the official eROSITA end-to-end simulator. SIXTE traces the photons through the optics by using the measured Point Spread Function (PSF) and vignetting curves \citep{predehl_erosita_2021} onto the detector. The input ideal photon list is convolved with the Redistribution Matrix File (RMF) and Auxiliary Response File (ARF) of the instrument. All 7 Telescope Modules (TMs) use the same RMF and a low-energy threshold of 60 eV. The detection process includes a detailed model of the charge cloud and read-out process. The software allows the independent processing of all 7 TMs: five of these (TMs 1, 2, 3, 4, and 6) have identical effective area curves, and TMs 5 and 7 have identical effective area curves, due to the absence of the aluminium on-chip optical light filter \citep{predehl_erosita_2021}. SIXTE can model the eROSITA's unvignetted background component due to high energy particles (hitting the camera and generating secondary X-ray emission) and electronic noise \citep{liu_establishing_2022}. We perform mock observations of eRASS:4 (in scanning mode) using the theoretical attitude file for the three components separately and then combine the event files.
\par
We need to include a physical background to LC30, given that the included physics is limited within $z<0.2$ and the X-ray cosmic background is dominated by the AGN population \citep{haardt_modelling_2001, gilli_synthesis_2007}. A realistic representation can be modelled with the deep background modelling extracted within the eFEDS area \citep[see][]{liu_erosita_2022}. The eROSITA background has been found to be relatively constant in time \citep{predehl_erosita_2021, brunner_erosita_2022, yeung_srgerosita_2023} and volume \citep{liu_establishing_2022}, we model the physical background as one single extended source with a flat surface brightness. In LC30, the simulated background in all 7 TMs is taken from \cite{liu_establishing_2022} and represents the spectral emission from all the unresolved sources in the eFEDS catalogue. 
\par
\begin{figure*}
    \centering
    \includegraphics[scale=0.38]{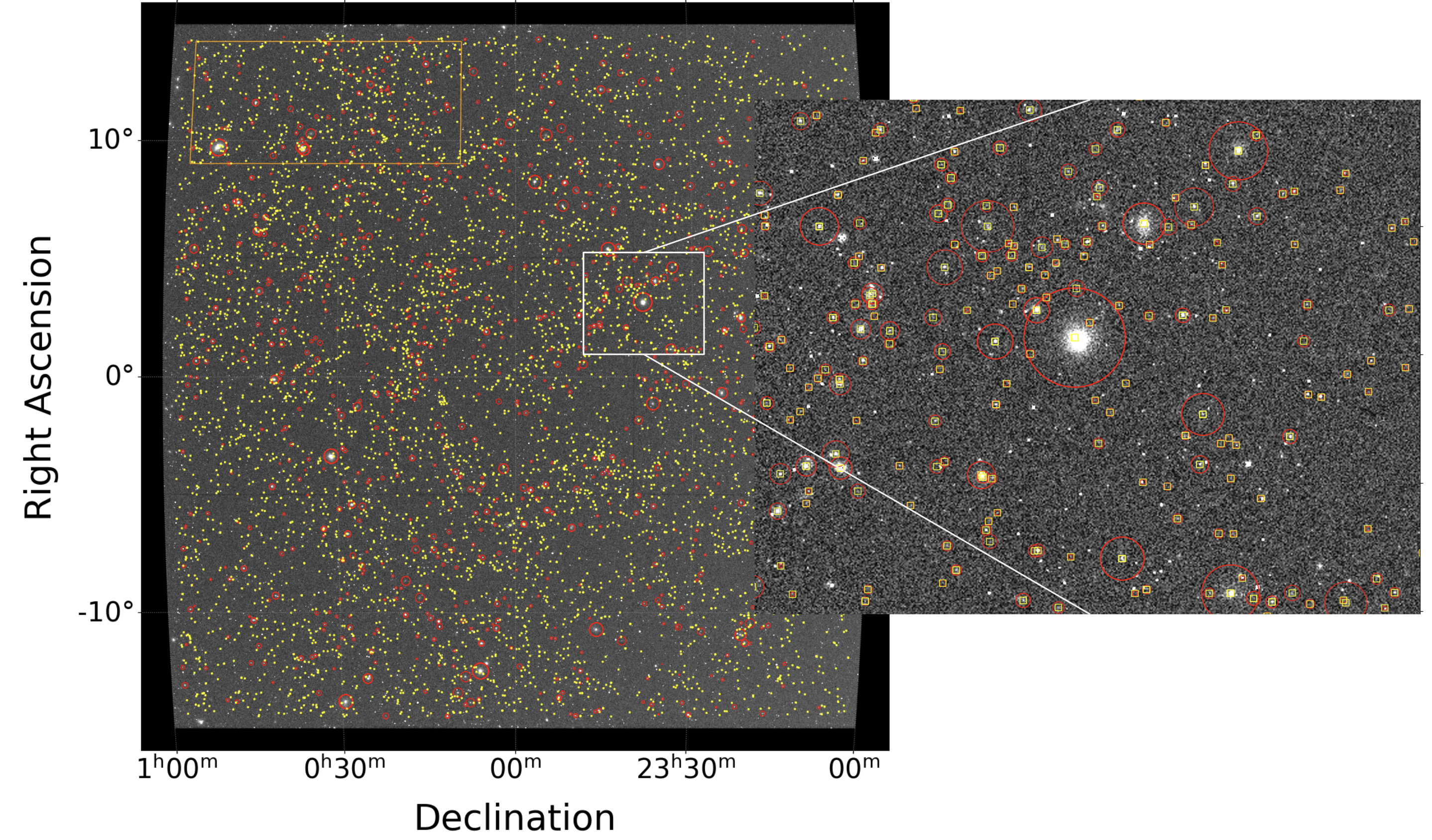}
    \caption{The FOV of LC30 includes all the X-ray sources modelled (i.e., IGrM, AGNs, XRBs and background) and a zoom-in over one of the largest clusters and surrounding environment at the lowest snapshot. We report in red the extended detections and with yellow squares the point-source detections. The circles mark the extension of the source being $R_{200}$ from the matched input catalogue. The orange rectangle at the top left shows the size of FOV for the eFEDS catalogue (i.e., 140 deg$^{2}$).}
    \label{fig:sky+regions+eFEDS+bkg}
\end{figure*}

Fig.~\ref{fig:sky+regions+eFEDS+bkg} illustrates the resulting mock X-ray Field of View (FOV) with the eSASS detections (see next section).  The full emission is included (i.e., IGrM, AGNs, XRB and background). Additionally, we report for comparison the eFEDS'FOV with an orange rectangle in the top-left corner of the image. We zoom in over one of the largest systems at the lowest redshift. 

\subsection{eSASS source detection}
\label{subsec:3.3}
The reduction of the simulated eROSITA data presented here follows the same data flow as the reduction of the eRASS data described in \cite{merloni_srgerosita_2024}. The event files of all emission components and all 7 eROSITA TMs are merged and filtered for photon energies within the $0.2-2.3$ keV band. The filtered events are binned into images with pixel size $4\arcsec$ and $3240 \times 3240$ pixels, corresponding to overlapping sky tiles with size $3.6 \times 3.6$ deg$^{2}$ having a unique area of $3.0 \times 3.0$ deg$^{2}$. The LC30 is covered by 122 of the standard eRASS sky tiles.
 
On each of these sky tiles the eSASS source detection chain described in  \citet{brunner_erosita_2022} is applied.
Based on the attitude file used for the SIXTE simulations, we calculate exposure maps and detection masks for each of the tiles using the eSASS tasks \texttt{expmap} and \texttt{ermask}.
The task \texttt{erbox} searches for significant excesses in the images and creates an initial source list.
After masking out the initial sources from the images, a background map is calculated using an
adaptive smoothing algorithm. This procedure is iterated three times using the resulting background maps
to refine the source search by \texttt{erbox}.
The final \texttt{erbox} lists and background maps are then passed to the PSF fitting task \texttt{ermldet}. This task models the 
PSF for each photon event based on its energy and detector position. At the input source positions
point source and extended source model are fitted to the data. The resulting source catalogue
contains the fitting parameters for positions, fluxes, and extent for each source. 
The calculated likelihood values for detection ($\mathcal{L}_{DET}$) and extent ($\mathcal{L}_{EXT}>0$) are related to the probabilities $P$ for spurious detection by
\begin{equation}
 \mathcal{L} = - \ln P\ \;  .  
\end{equation}
See Appendix A in \cite{brunner_erosita_2022} for further details.
\par
Finally, the \texttt{ermldet} source lists from the non-overlapping $3 \times 3$ deg$^{2}$ area of each sky tile are merged into the final source catalogue.
\section{The eSASS catalogue}
\label{sec:eSASS_catalogue}
In this section, we present the X-ray catalogue created with the eSASS pipeline on LC30 and the matching procedure. We also discuss the types of detections (primary and secondary), the integrated luminosity and the use of luminosity as mass-proxy through scaling relations.
\subsection{Extended sources: matching with the input catalogue}
\begin{figure}
    \centering
    \includegraphics[scale=0.6]{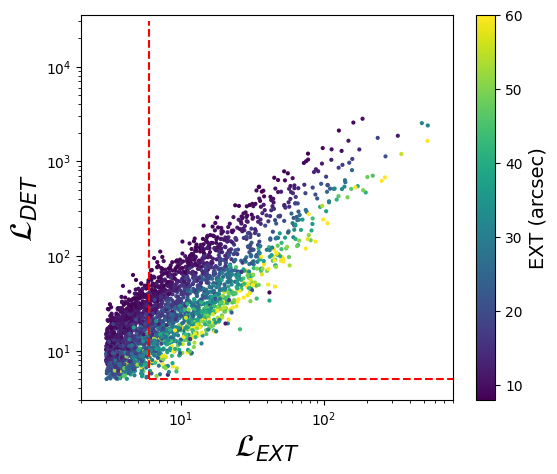}
    \caption{Distribution of the extended $\mathcal{L}_{EXT}$ and detection likelihoods $\mathcal{L}_{DET}$ for all the (candidate) extended sources. We colour-coded the points according to their angular scale in arcseconds. The red dashed lines mark the separation between the extent sources above the likelihood we choose to operate (i.e., $\mathcal{L}_{DET}>5$ and $\mathcal{L}_{EXT}>6$). }
    \label{fig:distribution_likelihood2}
\end{figure}

Similarly to eFEDS \citep{liu_2022groupsclusters} and the eRASS1 \citep{bulbul_srgerosita_2024} catalogues, we select the detections only in the soft band ($0.2-2.3$ keV). The total number of detections is 44459 whose vast majority (i.e., $\sim 94$\%) is composed of point sources. In Fig.~\ref{fig:distribution_likelihood2} we show the resulting distribution of all the 2674 extended detections ($\mathcal{L}_{EXT}>0$) as a function of the extent $\mathcal{L}_{EXT}$ and detection likelihoods $\mathcal{L}_{DET}$. We colour-code the points according to the corresponding angular extension in the catalogue. It is not surprising that at constant $\mathcal{L}_{DET}$, the higher $\mathcal{L}_{EXT}$ corresponds to the detections with the largest angular extension. In Fig.~\ref{fig:distribution_likelihood2} we also draw the minimum values (red dashed lines) for which $\mathcal{L}_{DET}$ and $\mathcal{L}_{EXT}$ will be selected as extended sources (i.e., 1675 candidates) in later analyses. 
\par
We evaluate the completeness and contamination of the catalogue by matching it with the input halo catalogue in the mock.
\par
The matching procedure goes as follows.
\begin{enumerate}
    \item For each extended source, we search for a halo counterpart in the {\it Magneticum} catalogue which lays within a maximum offset of $R_{500}$ from the detection's centre. Being the X-ray emissivity proportional to the electron density square, it is reasonable to direct our matching to the densest regions of the halos. 
    \item If more than one halo falls within the circular aperture, we check whether there is one whose flux dominates (i.e. at least four times the second brightest source). If this is the case, we match it, otherwise, we define a primary (the brightest detection) and a secondary source not detected due to blending. Even though there might be more than one very bright source in the aperture, we keep track only of the second brightest and flag the primary as blended. 
    \item At the end, we check whether according to these criteria, any halo in the input catalogue has been matched more than once to an eSASS detection. If so, we define the matched sources as fragmented. Fragmentation occurs in large halos that are split into smaller emissions by eSASS, thus we define the primary detection as the one with the highest flux while the fragments are labelled as secondary.
\end{enumerate}
Thus, according to the likelihood of the matching, we identify two types of detected sources: primary and secondary. Primary detections are either the unique or primary emitting extended source matched. Secondary extended detections are all sources that are impacted by blending or fragmentation. Our reference run is set for $\mathcal{L}_{DET}>5$ and $\mathcal{L}_{EXT}>6$, similarly to the eFEDS catalogue \citep{brunner_erosita_2022}. We run the unmatched catalogue through a second matching iteration with the eSASS catalogue of detected point sources (i.e, $\mathcal{L}_{DET}>5$ and $\mathcal{L}_{EXT}=0$) to account for the possibility that a fraction of these halos might be misclassified as point sources due to the sizable PSF of the detector \citep[e.g.,][]{xu_new_2018, bulbul_erosita_2022}. We limit our query to point detections whose flux is dominated (>50\%) by the IGrM photons. Going back to Fig.~\ref{fig:sky+regions+eFEDS+bkg}, we illustrate the detections in LC30: the red circles encompass the angular $R_{200}$ of the source, while the yellow squares correspond to the point-source detections.
\subsection{Extended sources: primary detections}
\begin{figure*}
        \centering
    \includegraphics[scale=0.5]{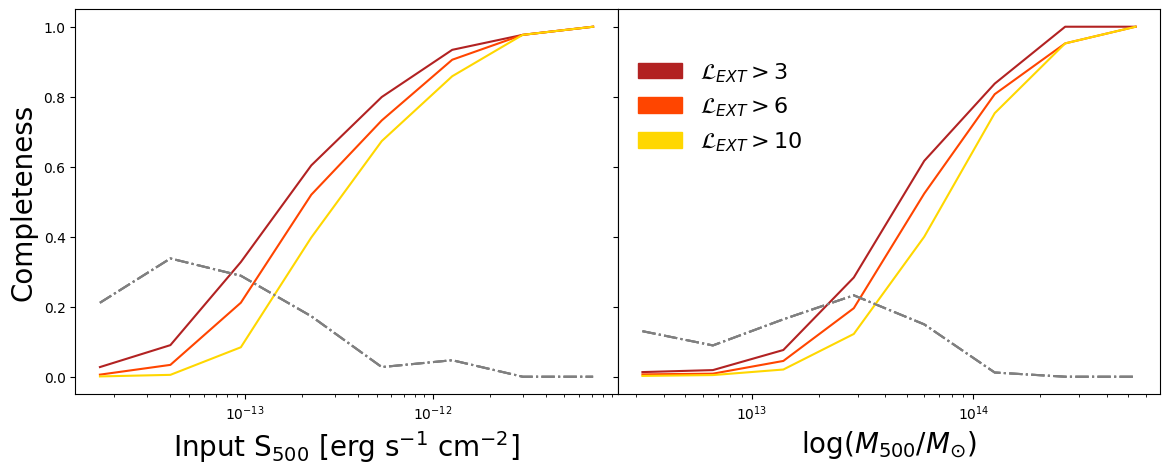}
    \caption{Completeness profiles within the three different extent likelihood cuts (i.e., $\mathcal{L}_{EXT}>3, 6, 10$). The completeness is defined in terms of the input flux $S_{500}$ in the soft band ($0.2-2.3$ keV) in the left panel and the halo mass $M_{500}$ in the right panel. We present the results for the sample with the extended (primary) detections with the solid lines. The dashed-dotted grey lines show the fraction of misclassified point sources: they are detections from the point source catalogue whose X-ray emission is at least 80\% from the IGrM in input.}
    \label{fig:completeness}
\end{figure*}

In this section, we check the completeness and contamination of the extended X-ray emissions to understand the detectability of the halo population. Although we perform the matching with the entire halo catalogue, we discuss only the detections corresponding to halos with mass $M_{200}\geq 10^{12.5}\, M_{\odot}$ (i.e., due to the simulation's resolution). In the following, we will refer to quantities with the subscript 
\begin{itemize}
    \item "EXT" referencing the extended detection subsample;
    \item "HALOS" for all the {\it Magneticum} groups and clusters;
    \item "SPUR" for the spurious sources (i.e., unmatched detections) in the eSASS catalogue;
    \item "eSASS" for all the detections following the eSASS pipeline.
\end{itemize}Therefore, we define completeness as the ratio between the fraction of detected halos over the total, namely:
\begin{equation}
    \mathcal{C_\mathrm{ompleteness}} = \frac{N_\mathrm{EXT}}{N_\mathrm{HALOS}}.
\end{equation}
We estimate this quantity (within $R_{500}$) as a function of the input flux $S_{500}$ and halo mass $M_{500}$ in the {\it Magneticum} catalogue. Similarly, we define contamination as the fraction of spurious detections in the eSASS catalogue:
\begin{equation}
    \mathcal{C_\mathrm{ontamination}} = \frac{N_\mathrm{SPUR}}{N_\mathrm{eSASS}}.
\end{equation}
Notice that this definition differs from the one provided in \cite{seppi_detecting_2022, liu_establishing_2022} and is equivalent to their spurious fraction. 
These two quantities are complementary in describing the detection process: completeness defines the accuracy of recovering the input catalogue whereas contamination clarifies to what extent it is susceptible to false detections. Such spurious detections can be due to the (random) background fluctuations that mimic source emission.
In Fig.~\ref{fig:completeness} we present the completeness of our X-ray extended catalogue within bins of the input flux within the eROSITA detection band (i.e., $0.2-2.3$ keV) in the left panel and halo mass in the right panel. 
\par
We perform the matching with the extended catalogue first and then the point source catalogue (with $>70\%$ emission from IGrM), as outlined in the previous section, however, we vary the cut on the extent likelihood within 3 values: above 3, 6 and 10. An extent likelihood $\mathcal{L}_{EXT}>6$ is used in constructing the eFEDS catalogue \citep{liu_2022groupsclusters}, thus it will represent our reference model throughout this analysis. On the other hand, an extent likelihood $\mathcal{L}_{EXT}>3$ is employed for the eRASS1 catalogue \citep{bulbul_srgerosita_2024} to maximize source discovery. Finally, we include the case for $\mathcal{L}_{EXT}>10$ to check how many of the spurious sources are left out in the process. Not surprisingly, there is a trend of increasing completeness with decreasing $\mathcal{L}_{EXT}$ threshold: a completeness of 90\% is reached at input flux $S_{500} = 1.23 \times 10^{-12}$ erg s$^{-1}$ cm$^{-2}$ (halo mass $M_{500} = 2.1\times 10^{14} M_{\odot} $) in the $\mathcal{L}_{EXT}>6$ sample. Changing the likelihood threshold allows us to gain only a few per cent on the flux and mass. However, lowering the extent likelihood threshold comes at the expense of the contamination, which increases resulting in  2\%, 3\% and 5\% for the $\mathcal{L}_{EXT}$ larger than 10, 6 and 3 respectively. We report with a dotted-dashed grey line the fraction of halos which are misclassified as point sources since their X-ray emission in input is at least 80\% coming from the IGrM. Most of these are small halos, with a peak at masses around $3\times 10^{13}\, M_{\odot}$.
\par
In the rest of the paper, our baseline model will be represented by the sample selected with the detection limits adopted for eFEDS, namely $\mathcal{L}_{EXT}>6$ and $\mathcal{L}_{DET}>5$.

\subsection{Extended sources: secondary detections}
This section aims to further discuss the class of secondary sources and their characteristics. Blended and fragmented sources fall in this class. Around 5\% of the primary detections in the eSASS catalogue ($\mathcal{L}_{EXT}>6$) include a second bright source which is not resolved. \cite{liu_erosita_2022} describe this to be the main reason for misclassification in their analysis. Source blending causes a secondary effect on the completeness of the data sample, as the effective observed sources will be reduced. In our case, blended sources impact only mildly the low-mass group regime $\log(M_{200}/M_{\odot}) \in [12.5, 13.5]$ an effect that cannot fully explain the drop in the number counts of low-mass systems in Fig.~\ref{fig:completeness}.
\par
Very bright and close-up sources can potentially be split into smaller detections by eSASS. We flag as fragmented all sources with more than one counterpart in the eSASS detections. In our sample, this is the largest source of misclassification, as around 12\% of the matched halos are fragmented into smaller halos by eSASS. Similarly to \cite{seppi_detecting_2022}, we discover that the fragmentation correlates mostly with the flux of the source, rather than its extension in the sky. In other words, sources with an extended emission will be split into smaller sources preferentially if their input flux is larger than $10^{-12}$ erg s$^{-1}$ cm$^{-2}$.
\par
In an observational analysis, secondary detections are hard to recover. If a source is fragmented into smaller detections, optical follow-up and aperture photometry allow to re-calculate the flux of the extended source. On the other hand, blended sources are confined to be counted as a single source throughout the analysis. Therefore, we decide to continue investigating these sources recalculating the flux within an aperture (see next section) and considering blended sources as a single source.  

\subsection{Extended sources: imaging and luminosity}
\label{sec:imaging&luminosity}
To mitigate the source fragmentation effect and to closely mimic the observational approach, we recompute the rest-frame soft band luminosity of all the extended sources in an aperture equal to $R_{500}$ of the input halo, rather than using the input flux extracted from the {\it Magneticum} simulation.
The imaging analysis in this work is based on a direct image fitting by deconvolving the PSF and converting the de-projected count rate profile into a surface brightness profile assuming an energy conversion factor (ECF). The ECF is computed through XSPEC assuming an absorbed \texttt{apec} model with an emission measure of $1$. We assume a gas temperature extracted via the $M_{500}-T_{500}$ scaling relation in \cite{lovisari_scaling_2015} and a fixed metallicity $Z=0.3\, Z_{\odot}$ \citep[like in \texttt{PHOX}][]{anders_abundances_1989}. We use pyproffit \citep{eckert_low-scatter_2020} for the task and restrict to the soft energy band ($0.5-2.0$ keV). Images and exposure maps (vignetted and unvignetted) were extracted using the eSASS tools \texttt{evtool} and \texttt{expmap}. 
\par
\begin{figure}
    \centering
    \includegraphics[scale=0.5]{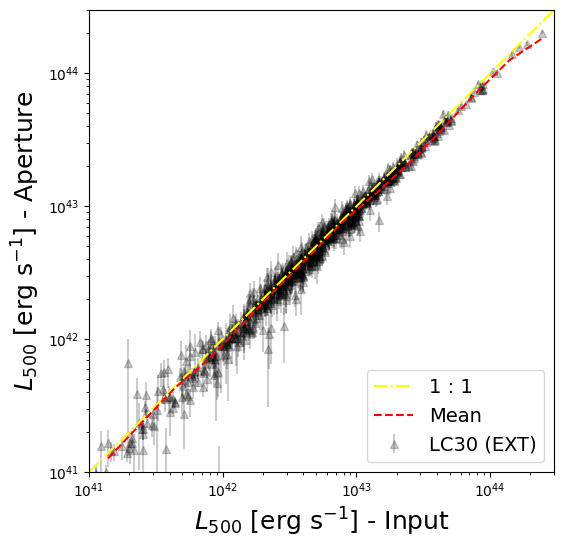}
    \caption{Comparison between the soft band luminosity recovered from flux aperture in the mock observation and the input luminosity of the halos in the lighcone. The uncertainties are calculated as the 16$^{th}$ and 84$^{th}$ percentile from the MCMC run. We overplot the 1:1 relation in yellow and the best-fit relation in red.}
    \label{fig:LumVSLum}
\end{figure}

We compare these rest-frame luminosities and their associated errors (16$^{th}-$84$^{th}$ percentile from the MCMC in pyproffit) with the intrinsic luminosity of the sources directly calculated from the mock in Fig.~\ref{fig:LumVSLum}. Additionally, we plot in the figure the 1:1 curve in yellow and the running mean in red, to highlight the offset we find between the two estimates. The re-calculated luminosities are mostly consistent with the input luminosities, although slightly underestimated at the high-luminosity end. The scatter in the relation seems to be dependent on the input luminosity, increasing towards fainter halos reaching 20\% at maximum in the faintest end. This is not surprising as the recovered luminosities also exhibit larger relative uncertainties. In this regime, the number of events measured in the chosen aperture is largely reduced due to the intrinsic faintness of the IGrM: colder gas, metallicity and lower gas fractions dramatically impact the X-ray emissivity \citep{eckert_feedback_2021, oppenheimer_simulating_2021}.
\par
Additionally, Galactic absorption further damps the emission of such systems. At the high-luminosity end, the impact of assuming a single temperature to estimate to fit the surface brightness profile and constrain the luminosity spans $\sim 5$\% \citep{zuhone_effects_2023}, much smaller than the expected statistical uncertainty from the eROSITA observations.
\par
\begin{figure}
    \centering
    \includegraphics[scale=0.7]{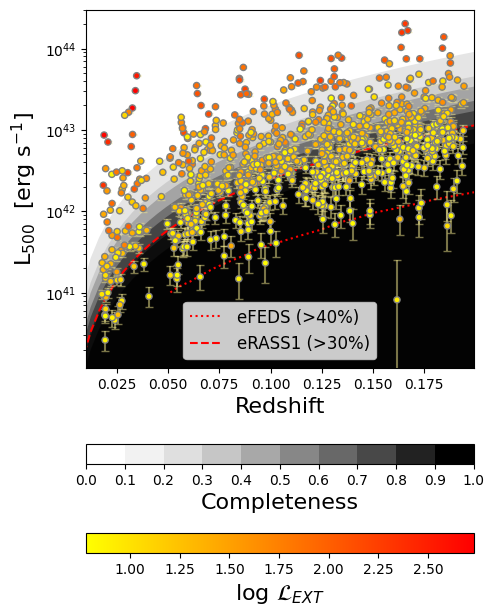}
    \caption{Soft band luminosity distribution of the groups and clusters in the lightcone in function of their redshift. The errors are the 16$^{th}$ and 84$^{th}$ percentile from the MCMC process to extract the luminosity. The data points are colour-coded for their extent likelihood. The shading areas represent the range of survey flux limits given by the completeness from 10\% to 90\% in steps of 10\% difference with decreasing colour. The red curves represent the flux limit in eFEDS \citep[$1.5 \times 10^{-14}$ erg~s$^{-1}$~cm$^{-2}$;][]{liu_2022groupsclusters} and eRASS1 \citep[$10^{-13}$ erg~s$^{-1}$~cm$^{-2}$][]{seppi_detecting_2022} for 40 and 30\% completeness respectively. }
    \label{fig:luminosity_z}
\end{figure}

In Fig.~\ref{fig:luminosity_z}, we show the distribution of the luminosity as a function of the redshift of the source. The error bars represent the 16$^{th}$ and 84$^{th}$ percentile from the Monte Carlo Markov Chain (MCMC) run in pyproffit. We colour-code the points according to their extent likelihood in the eSASS catalogue in arcseconds. We overplot shaded areas representing the survey flux limit at different rates of completeness: from 10\% (corresponding to $4.5 \times 10^{-14}$ erg~s$^{-1}$~cm$^{-2}$) in light grey to 90\% ($8 \times 10^{-13}$ erg~s$^{-1}$~cm$^{-2}$) with dark grey in steps of 10\% difference. We notice that the relation goes through an underdense region between $0.035\leq z \leq 0.05$ which corresponds to a small void in the lightcone. The red curves represent the flux limits for 40\% and 30\% completeness for the eFEDS and eRASS1 surveys respectively.
\subsection{Extended sources: mass recovery}
The matching technique allows us to infer the intrinsic properties of the detected and undetected halos and gather information regarding the biases and systematics which affect the detection process \citep{bulbul_erosita_2022, popesso_x-ray_2023}. Future works will further unveil the effects on scaling relations (Toptun et al. in prep, Dev et al. in prep). Here, we discuss the consequences of using the estimated intrinsic luminosity to recover the X-ray observables, such as the halo mass $M_{500}$.
\par
Together with redshift, the halo mass allows us to test models of structure formation and constrain cosmological parameters \citep{grandis_impact_2019, pratt_galaxy_2019}. The most common and reliable mass-proxies are calibrated through galaxy kinematics \citep{pratt_galaxy_2019}, X-ray temperature via hydrostatic equilibrium and weak lensing \citep{umetsu_cluster-galaxy_2020}. When such methods are unavailable, one has to rely on other estimators. Assuming that the hot intracluster medium follows the scale-free gravitational collapse of DM, \cite{kaiser_evolution_1986} derive X-ray scaling relations among different thermodynamic properties. The calibration of the X-ray $L_{500}-M_{500}$ relation is extensively discussed in the literature highlighting that most findings exhibit slopes from $1.4$ to $1.9$ steeper due to the presence of non-gravitational processes \citep[][for a review]{lovisari_scaling_2021}. Therefore we want to test the systematics and uncertainties derived by using solely the X-ray luminosity, a scenario similar to that of shallow X-ray surveys.
\par
For the experiment, we calculate the halo masses through the scaling relation derived in \cite{lovisari_scaling_2015} which is based on a sample of $\sim 80$ between clusters and groups of galaxies, whose masses have been determined by X-ray data and hydrostatic equilibrium assumption. Our choice stands with the scaling relation used to determine the radius $R_{500}$ (and thus, the mass $M_{500}$) in the eFEDS analysis \citep{liu_2022groupsclusters} which also results as the closest relation to our best-fit. As a second test, we employ the scaling relation deduced with a subsample from the eFEDS clusters and groups catalogue \citep{chiu_erosita_2022} using weak lensing measurement covered by the Hyper Suprime-Cam \citep{aihara_hyper_2018}. Despite no significant difference between the luminosity and temperature of the detected clusters in eRASS1 and eFEDS \citep{bulbul_srgerosita_2024}, some discrepancies have been reported in clusters and groups from eRASS1 between luminosities \citep[eROSITA and Chandra,][]{bulbul_srgerosita_2024}) and temperatures \citep[eROSITA and XMM-Netwon][]{migkas_srgerosita_2024} measured with different telescopes. Thus, we expect the scaling relation $L_{500}-M_{500}$ derived in \cite{chiu_erosita_2022} to differ from others. 
\par
\begin{figure}
    \centering
    \includegraphics[scale=0.5]{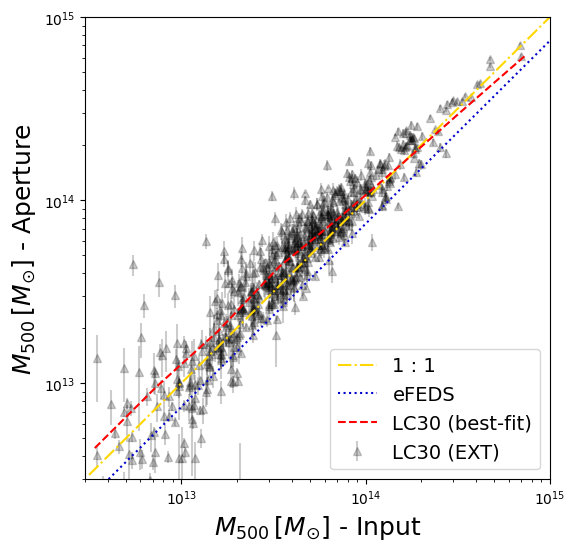}
    \caption{Comparison of the mass distribution in the input catalogue and derived from the scaling relation $L_{500}-M_{500}$ in \cite{lovisari_scaling_2015}. The errors are derived from the 16$^{th}$ and 84$^{th}$ percentile in the luminosity measurement. In dash-dotted yellow, we report the 1:1 relation; in dashed red the best fit from the scatter plot. Additionally, we determine the mass using the scaling relation derived from eFEDS data in \cite{chiu_erosita_2022} and plot its best fit as dotted blue.  }
    \label{fig:MassVSMass}
\end{figure}

Fig.~\ref{fig:MassVSMass} reports our findings. The distribution of points marks the estimated masses for the single halos when applying the scaling relation in \cite{lovisari_scaling_2015}. We notice that this relation is calibrated for an X-ray luminosity in the $0.1-2.4$ keV energy band, whereas our luminosity is estimated in the band $0.5-2.0$ keV, therefore we correct the normalization assuming a conversion factor of  $1.34$ (assuming an absorbed \texttt{apec} emission with a mean redshift of $0.1$, temperature $1$ keV, abundance $0.3$ and Galactic absorption with column density of $N_H=10^{20}$ cm$^{-2}$). Furthermore, we check with the input luminosity in this band the accuracy of such prescription and the lack of systematics at this stage.
In dash-dotted yellow, we overplot the 1:1 relation to compare the dashed red, namely the best-fit of the points shown -- whose masses are calculated with \cite{lovisari_scaling_2015} -- and the dotted blue corresponding to the best-fit of the points (not shown) whose masses are calibrated with \cite{chiu_erosita_2022}. Although we do not discover a constant offset in the mass with the former, we note significant differences in the distribution. At the low-mass end, the distribution presents large relative uncertainties and a large scatter from the 1:1 relation: the scatter in the scaling relation is around $\sim 20\%$ at fixed mass \citep{lovisari_scaling_2015}. On the other hand, the mass estimated with the scaling relation from eFEDS \citep{bahar_erosita_2022} provides masses 20\% larger in the whole range, as illustrated by the best fit in the figure. We do not attempt to determine the tilt in the eFEDS sample. 
\par
Among the many uncertainties on the observational side, we argue that an accurate estimate of the X-ray emission within the halo radius (e.g., $R_{500}$) can moderately impact the luminosity, and in turn the scaling relation. In Appendix \ref{sec:appendix}, we provide a brief discussion on the impact of the flux estimates when the radius is not properly estimated. In our simulated run, we expect such deviations to happen mostly at the high mass end, where the surface brightness profile is steeper. Furthermore, any deviation from the condition of hydrostatic equilibrium can bias the estimates of cluster masses \citep[10-20\%]{rasia_systematics_2006, nagai_testing_2007,pratt_galaxy_2009, biffi_nature_2016, gianfagna_exploring_2021}. Cosmological accretion and turbulent motion \citep{vazza_turbulence_2017, gaspari_linking_2020}, viscosity and thermal conductivity \citep{zuhone_effect_2015,gaspari_chaotic_2015} can all provide such circumstances. Additionally, \cite{zuhone_effects_2023} show that projection effects have a negligible impact. 
\par
We argue that the uncertainty in the mass estimation derived from the mass-luminosity scaling relation is unbiased but carries a significant scatter \citep[$\sim 20-30\%$][]{lovisari_scaling_2015}. The possibility of using an alternative mass estimator that exploits the tightness of the mass-temperature relation would considerably decrease the uncertainty \citep[to $8-15$\%][]{rasia_temperature_2014} even though masses are usually deduced after a single temperature fit to the spectrum which tends to underestimate the masses of 10-15\% \citep{rasia_temperature_2014}.

\subsection{Point-sources: matching and luminosity}
Up to this point, we have treated point sources (i.e., $\mathcal{L}_\mathrm{EXT}=0$) with the sole purpose of re-discovering misclassified halos in the catalogue. In this section, we aim to further investigate their presence as pure detections in the eSASS catalogue, since they make up for 86\% of all the sources with detection likelihood $\mathcal{L}_\mathrm{DET}>10$. They are most predominantly composed of bright AGNs \citep{brandt_cosmic_2015} whose population is responsible for most of the cosmic (soft) X-ray background in the Universe \citep[e.g.,][]{schmidt_space_1968, gilli_synthesis_2007}. 
\par
We select point sources with $\mathcal{L}_\mathrm{DET}>10$, which should yield a completeness of 94\% in an eFEDS-like survey \citep{liu_erosita_2022}. We match them to our AGN input catalogue within a distance of 20" between the detection and the AGN position, similarly to \cite{seppi_detecting_2022}. We find a completeness of 93\% within the flux limit of $1.6 \times 10^{-14}$ erg~s$^{-1}$~cm$^{-2}$, in line with the findings in \cite{merloni_erosita_2012}. We include the possibility of blending between point sources, although only 2\% of all detections seem to suffer from it. 
\par
We extract the rest-frame soft band luminosity directly from the PSF fitting performed by eSASS assuming an energy conversion factor (ECF) between counts and flux in $0.5-2.0$ keV of $1.604\times 10^{12}$ (erg~cm$^{-2}$)$^{-1}$ \citep{brunner_erosita_2022}. The ECF is calculated assuming an absorbed power-law with $\Gamma=1.8$ and Galactic absorbing column density of $N_H=10^{20}$ cm$^{-2}$ which reflects our AGN modelling \citep{biffi_agn_2018}.

\section{The X-ray luminosity function}
\label{sec:luminosity_function}
The X-ray luminosity function (XLF) describes the distribution (i.e., number density) of X-ray sources in a survey as a function of their luminosity. The common approach to derive a non-parametric XLF is based on the $1/V_\mathrm{max}$ technique described in the seminal work by \cite{schmidt_space_1968} and later generalised by \cite{avni_simultaneous_1980}. Hence, the sources are parsed into luminosity bins $\Delta L$ such that in the \textit{i-th} bin the XLF centered on $L_{i}$ is 
\begin{equation}
    \frac{dn}{dL} (L_i) = \frac{1}{\Delta L_i} \sum_j \frac{1}{V_\mathrm{max}[L_j, F_\mathrm{lim}, A(F_j)]}.
\end{equation}
In the equation above, $V_\mathrm{max}$ is the (shell) comoving volume in which a halo \textit{j-th} of luminosity $L_{j}$ in the \textit{i-th} bin can be detected above the flux limit $F_\mathrm{lim}$ and sky coverage of the survey $A(F_j)$ (i.e., field of view of the lightcone: 900 deg$^2$). We estimate $L_i$ as the median luminosity of the halos in the bin. This prescription is similar to the one outlined in \cite{liu_2022groupsclusters} although in our approach we omit to include the selection function term. 

\subsection{Extended detections: groups and clusters}
\begin{figure}
    \centering
    \includegraphics[scale=0.55]{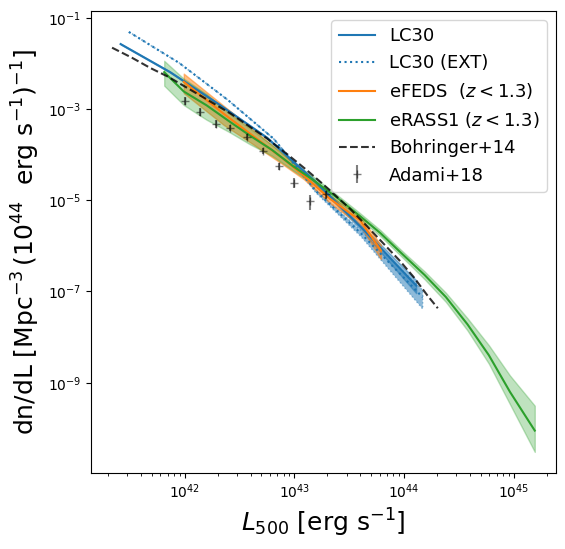}
    \caption{XLF of intrinsic soft band ($0.5-2.0$ keV) for the extended sources in the lightcone. We plot the results for the full sample (solid blue line) and for only detected ones (dotted blue line). The shaded bands represent the Poisson uncertainty in the luminosity bin. Similarly, we show the XLF for the eFEDS \citep{liu_2022groupsclusters} and eRASS1 \citep{bulbul_srgerosita_2024} data in orange and green respectively. For reference, we also add other results from the literature in black: the XXL-C1 sample \citep{adami_xxl_2018} and the REFLEX~II \citep{bohringer_extended_2014}.  }
    \label{fig:dndL_x}
\end{figure}

We determine the XLF for the extended (primary) sources in the catalogue and evaluate the differences by including all the X-ray sources in the lightcone. Even if eRASS:4 is not a flux-limited survey, we set the flux limit to $1.4\times 10^{-13}$~erg~s$^{-1}$~cm$^{-2}$ which corresponds to 50\% catalogue completeness (see Fig.~\ref{fig:luminosity_z}) and determines the upper bound to the redshift integration in the shell volume. In Fig.~\ref{fig:dndL_x} we show the lightcone curves in blue: the solid line represents the XLF calculated with input simulation luminosity for all the halos whereas the dotted line is the XLF estimated with the extended (primary) detections corrected for the effective volume. The correction for the effective volume $V_\mathrm{max}$ leads to a mildly steeper XLF than the theoretical one calculated from all the halos. The errors for the XLF given in the figure are the Poisson uncertainties on the number of objects per bin. We plot the XLF estimated using the eFEDS (in orange) and eRASS1 (in green) sample: these curves are normalised for the detection probability as described in \cite{liu_2022groupsclusters}. We remind that both curves are extracted for group and cluster samples up to $z=1.3$. We also report findings from the XXL-C1 sample in the $[0.01, 0.35]$ redshift range \citep{adami_xxl_2018} and from the REFLEX~II sample \citep{bohringer_extended_2014} from halos in the local Universe, rescaled to measure luminosities in the $0.5-2.0$ keV energy range. We find relatively good agreement at all luminosities among all sets, except with \cite{adami_xxl_2018} at $10^{43}$ erg~s$^{-1}$ which also has the largest uncertainties in that range. The lightcone XLFs seem to be mildly steeper than the observational counterpart, although still within 1$\sigma$.

\subsection{Point source detections: AGNs}
\begin{figure}
    \centering
    \includegraphics[scale=0.55]{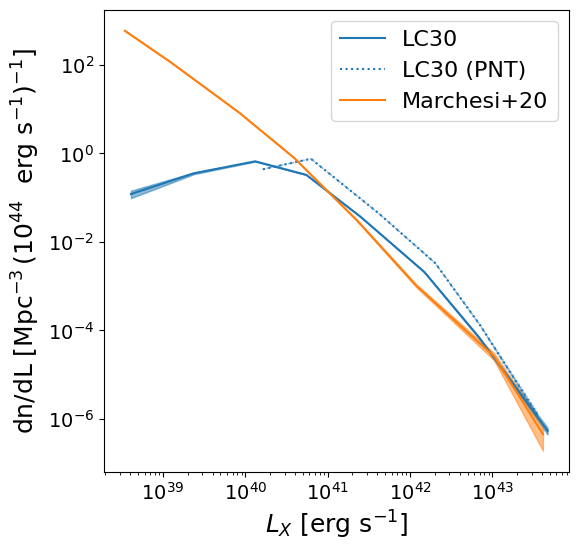}
    \caption{XLF of the intrinsic soft band ($0.5-2.0$ keV) for the AGNs in the lightcone. The luminosities are not corrected for Galactic absorption. We report the results for both the detected population (dotted blue line) and the entire population (solid blue line). In orange, we plot the XLF extracted from the mock catalogue in \cite{marchesi_mock_2020} properly rescaled to account for the survey volume difference. }
    \label{fig:dndL_x_AGN_eSASS}
\end{figure}

We extract the XLF for both detected and undetected AGNs within the lightcone. The validity of the XLF for bolometric luminosity has already been established in the {\it Magneticum} boxes across both high \citep{hirschmann_cosmological_2014,steinborn_refined_2015} and low redshifts \citep{biffi_agn_2018}, and \cite{biffi_agn_2018} have further corroborated its reliability for unabsorbed AGN emissions in both the soft and hard bands. Here, we test the XLF in the lightcone regime with Galactic absorption, thus we do not attempt to correct the fluxes to recover the unabsorbed emission. Fig.~\ref{fig:dndL_x_AGN_eSASS} illustrates the results in comparison with the XLF extracted from the mock catalogue presented in \cite{marchesi_mock_2020} restricted within a redshift range $[0.0, 0.2]$ and rescaled for the sky area (100 deg$^2$). We report the (intrinsic $0.5-2.0$ keV) XLF both from our detected population (dotted line) and the total population (solid line) of AGN. We find good agreement in the high luminosity regime ($\gtrsim 5 \times 10^{42}$ erg~s$^{-1}$), despite an overestimation at intermediate luminosities $\sim 10^{41} - 10^{42}$ erg~s$^{-1}$. The set from \cite{marchesi_mock_2020} is complete down to much lower fluxes ($5 \times 10^{-20}$ erg~s$^{-1}$~cm$^{-2}$), which explains the different shapes in the profiles at luminosities $<10^{41}$ erg~s$^{-1}$. At $L_{X}\sim5\times 10^{41}$ erg~s$^{-1}$, the completeness of the AGN catalogue is strongly reduced.
\par
The unresolved faint point sources will constitute part of the physical X-ray background in the mock mostly modelled by the real eFEDS background spectrum (see Sect.~\ref{sec:X-ray_catalogue}). Another fraction of this background is made up of the electronic noise modelled by SIXTE. 
\section{The hidden bias: undetected sources}
\label{sec:undetected}
Fig~\ref{fig:completeness} indicates that for masses below $M_{500}\leq10^{14}\, M_{\odot}$, the eRASS:4 completeness should drop quickly and be about 50\% at $10^{13.5}\, M_{\odot}$. In this section, we explore with the help of \textit{Magneticum} which gas properties might be the key driving the observed selection function. This is fundamental to understanding which group population the eRASS:4 selection function will be able to detect and identify the possible biases. We investigate the role of the electron density, and thus, the hot gas mass fraction, temperature and metallicity with the halo X-ray emissivity and thus, the detectability through eROSITA.

\subsection{What affects the X-ray emissivity?}
\begin{figure*}
    \centering
    \includegraphics[scale=0.6]{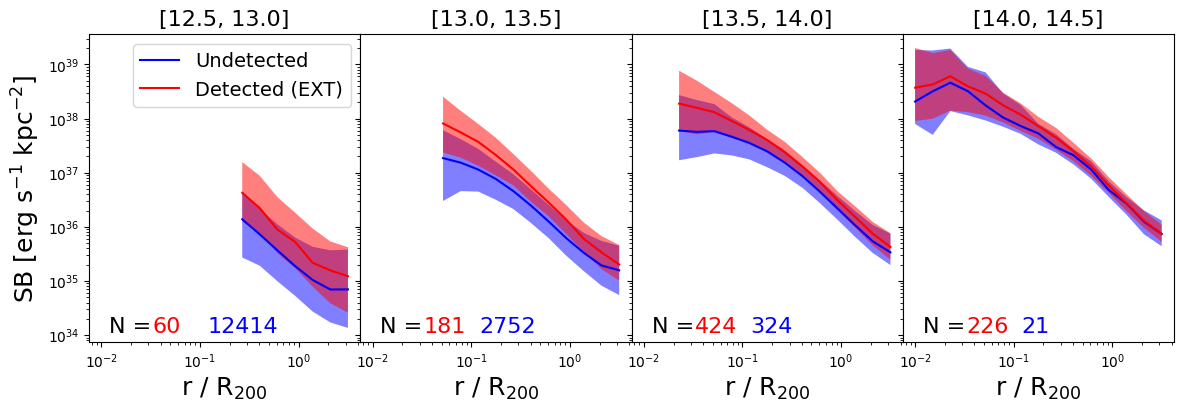}
    \caption{The median X-ray surface brightness of the extended detected (red) and undetected (blue) sources in the $0.5-2.0$ keV for four different logarithmic halo mass bins (i.e., $\log(M_{200}/M_{\odot})$ reported at the top of each panel). In the lower left of each panel, we report the number of detected and undetected in the mass bin. The shaded bands illustrate the 16$^{th}-$ 84$^{th}$ percentile from the intrinsic distribution. }
    \label{fig:SB}
\end{figure*}

Fig.~\ref{fig:SB} shows the resulting X-ray (rest-frame $0.5-2.0$ keV) surface brightness profiles from the extended detections and the undetected population in different halo mass bins (see top of each panel for the corresponding logarithmic $M_{200}$). In each panel, we also report the number of detected and undetected sources for each bin. We only include photons from the IGrM (thus, excluding background, AGN and XRB contributions) and exclude the inner core of the profiles when the softening length has similar scales. We find that the profiles are consistently shifted from one to another for all mass bins: the undetected population tends to have lower surface brightness making it harder to be detected by eSASS. These results are consistent with the studies reported in \cite{popesso_x-ray_2023}, where the authors show the average X-ray surface brightness profiles of the undetected galaxy group population when stacked on the optical peak emission derived from the GAMA survey \cite{driver_galaxy_2022}. \cite{popesso_x-ray_2023} argue that the undetected population may be the descendant of a post-merger phase resulting in the observed lower concentration and younger stellar populations. Such events could be linked to the location of the halo in the large-scale structure (e.g., void, filament) and/or the different mass accretion rates. 
\par
\begin{figure}
    \centering
    \includegraphics[scale=0.55]{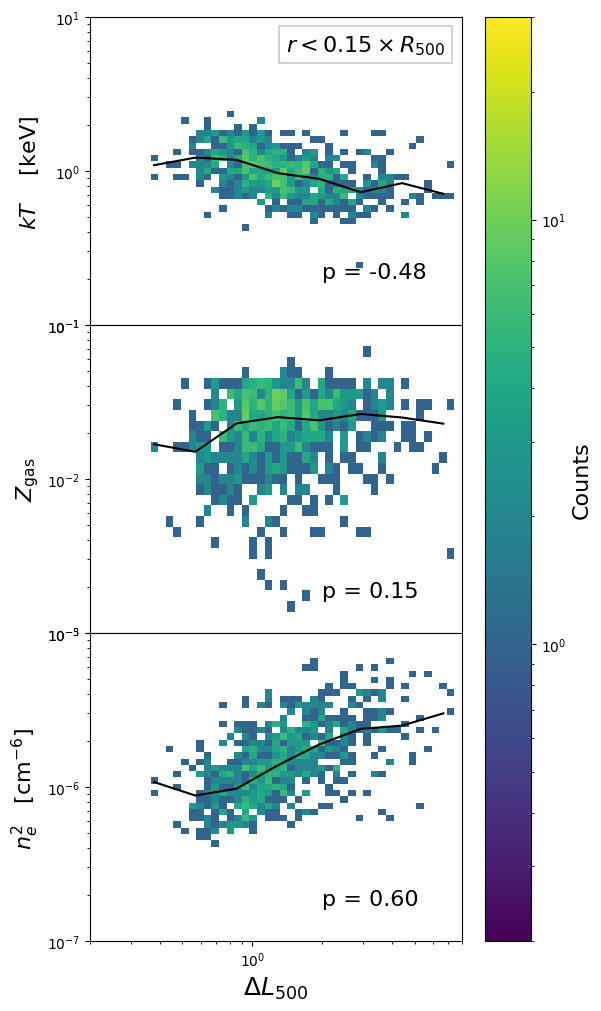}
    \caption{Properties of the gas as a function of the scatter in the scaling relation $L_{500}-M_{500}$ by \cite{lovisari_scaling_2015} in one halo mass bin ($10^{13}-10^{13.5}\, M_{\odot}$). From the top to bottom panels, we show temperature, metallicity and gas fraction (all mass-weighted) measured within $0.15 R_{500}$ of each halo. The black line denotes the running median for each quantity. We also report the Spearman coefficient in the plot.  }
    \label{fig:T_Z_Fgas_13-13.5}
\end{figure}

To investigate the physical reasons behind fainter surface brightness at fixed halo mass, we split the full sample as a function of equally log-spaced bins of halo mass $M_{200}$ and we characterise the X-ray emissivity (i.e., the energy emitted per time and volume), which reads
\begin{equation}
\label{eq:emissivity}
    \epsilon = n_e^2 \Lambda_{\nu}(T, Z).
\end{equation}
The electron density $n_e$ and the cooling function $\Lambda_{\nu}$ at frequency $\nu$ appear in the equation above and depend on the gas properties. Therefore, we estimate the dependence of the scatter of the $L_{500}-M_{500}$ relation of Fig.~\ref{fig:scaling_relations} with $n_e$, $T$ and $Z$. The scatter is estimated as residuals ($\Delta L_{500}$) with respect to the \cite{lovisari_scaling_2015} relation, which best represents the {\it Magneticum} data (see Fig.~\ref{fig:scaling_relations}).
The analysis is performed within different radii from $0.15 R_{500}$ to $R_{500}$. We restrict to halos with at least 80 particles within the selected sphere to provide a robust statistical measurement and in the halo mass bin $10^{13}-10^{13.5}\, M_{\odot}$. To guide the eye, we also overplot the running median as a function of the scatter with a black line. We find the highest significance dependence within the system core at $0.15 R_{500}$. Fig.~\ref{fig:T_Z_Fgas_13-13.5} shows the residuals $\Delta L_{500}$ versus the three gas properties in a single halo mass bin. The same trends are found at all halo masses.
\par
The Spearman coefficients for the $n_e^2$, $T$ and $Z$ are -0.48, 0.15 and 0.60, respectively, indicating that the strongest dependence comes from the electron density. A poorly significant anticorrelation is found with the temperature, while no dependence is found from the metallicity. The Spearman coefficient decreases by 35\% going from $0.15 R_{500}$ to $R_{500}$. This is consistent with the fact that the emission from the core region is dominating the emissivity and the system detectability from eROSITA as already found in \cite{clerc_synthetic_2018, seppi_detecting_2022}. The lower $n_e^2$, the lower the X-ray luminosity at fixed halo mass. This points to a dependence on the hot gas mass fraction suggesting low $f_\mathrm{gas}$ halos are under-luminous in the X-rays at fixed mass.
\par
\begin{figure}
    \centering
    \includegraphics[scale=0.5]{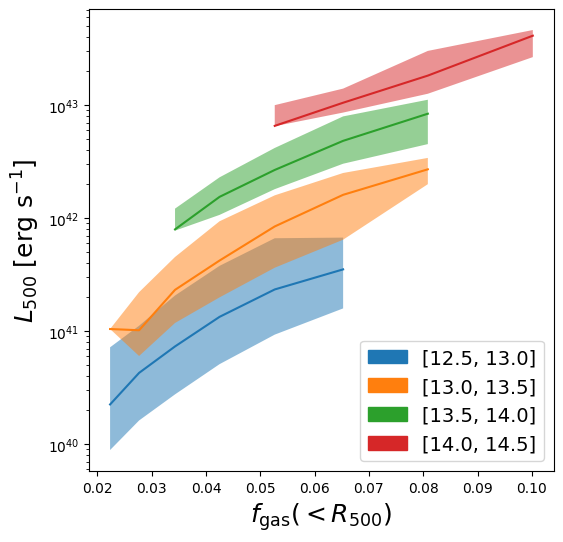}
    \caption{Luminosity of the IGrM as a function of the fraction of hot gas within the same radius (i.e., $R_{500}$). The halos are binned in logarithmic halo mass $M_{200}$ (see legend). The shaded bands mark the 68$^{th}$ confidence intervals.}
    \label{fig:L500_Fgas}
\end{figure}

We generalise this discussion in Fig.~\ref{fig:L500_Fgas} which illustrates the effect of the (mass-weighted) hot gas fraction in the X-ray luminosity ($0.5-2.0$ keV) integrated up to $R_{500}$: lower gas fractions correspond to lower luminosities. Therefore, the faint undetected population must retain on average lower hot gas fractions than the detected halos.  These findings were already discussed in the cluster regime in \cite{ragagnin_simulation_2022} for the \textit{Magneticum} simulation, but we see here that this result holds in the group regime as well.
\begin{figure}
    \centering
    \includegraphics[scale=0.6]{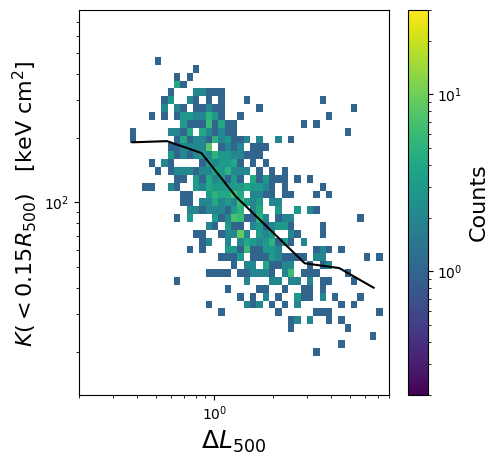}
    \caption{Entropy of the hot IGrM measured within $0.15R_{500}$. We plot it as a function of the scatter in the scaling relation $L_{500}-M_{500}$ by \cite{lovisari_scaling_2015} in one halo mass bin ($10^{13}-10^{13.5}\, M_{\odot}$).}
    \label{fig:entropy_deltaL500}
\end{figure}
\subsection{The central role of entropy}
It is of particular interest to further support this argument by showing the resulting entropy in these systems. The entropy $K$ of the IGrM is defined
\begin{equation}
    K = \frac{T}{n_e^{2/3}}
\end{equation}
where $T$ is the temperature of the gas and $n_e$ is the electron density. In Fig.~\ref{fig:entropy_deltaL500}, we present the entropy as a function of the scatter $\Delta L_{500}$ and we observe that indeed low-entropy systems (i.e., where the gas fraction at the centre is high) have the highest scatter $\Delta L_{500}$. Turning the argument around, we can say that the brightest sources correspond to sources with the lowest entropy intake at the centre. Such structural differences highlight a selection effect in our measurements whereby we favour samples at lower entropy (i.e., gas-rich).
\par
\begin{figure*}
    \centering
    \subfigure[]{\includegraphics[width=0.44\textwidth]{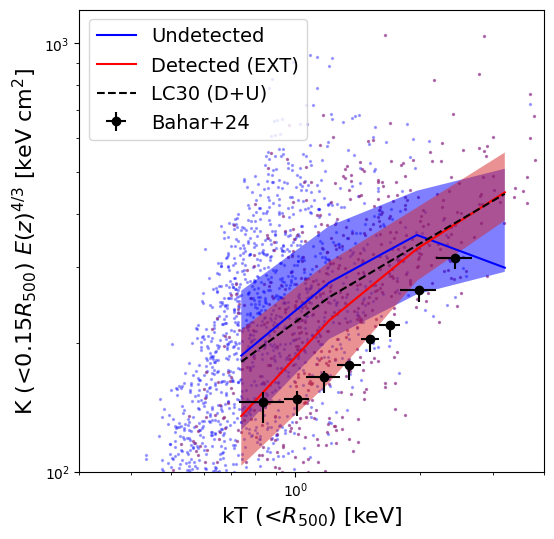}} 
    \subfigure[]{\includegraphics[width=0.44\textwidth]{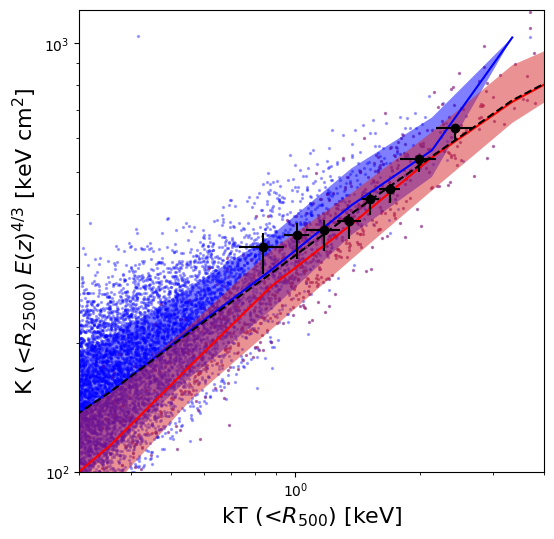}} 
    \subfigure[]{\includegraphics[width=0.44\textwidth]{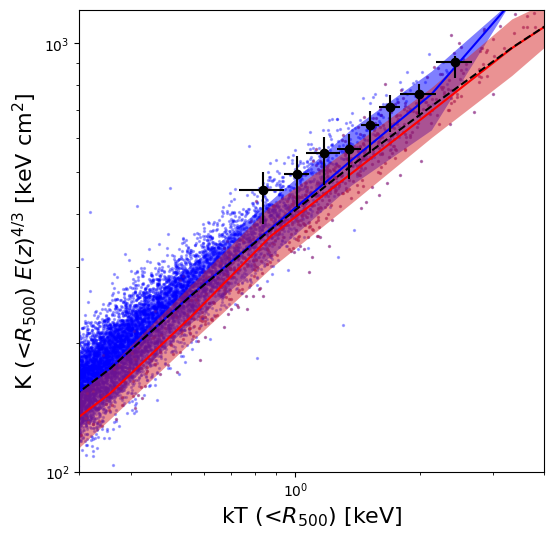}}
    \caption{Entropy within different radii as a function of the integrated temperature (within $R_{500}$). The entropy is calculated for each halo as the ratio between the mass-weighted temperature and the electron density within that same radius. We calculate the running median for the detected (red), undetected (blue) and full sample (black). We plot the $25^{th}-75^{th}$ percentile with the shaded band. The black crosses represent the observational eRASS1 dataset from \cite{bahar_srgerosita_2024}. The three images refer to three different radii (i.e., $0.15R_{500},\, R_{2500}, R_{500}$). We ensure halos host at least 80 gas particles within the selected aperture.}
    \label{fig:entropy}
\end{figure*}

This intrinsic bias can be highlighted by comparing the recent observational results published in \cite{bahar_srgerosita_2024} with our detected and undetected halos in Fig.~\ref{fig:entropy}. Following the approach adopted in \cite{bahar_srgerosita_2024}, we calculate the entropy averaged within three different radii (i.e., $0.15 R_{500}, R_{2500}$ and $R_{500}$) to illustrate the impact of the gas phase-space structure at different apertures. We remind \cite{bahar_srgerosita_2024} provides the results for the detections from the eRASS1 catalogue, although calculating the gas properties with the (deeper) eRASS:4 data. On the other hand, we mock an eRASS:4 lightcone with detections extracted in the eRASS:4 exposure time. This can potentially lead to small differences in the final results. Furthermore, we point out that, although comparable, the profiles we extract for the full sample of halos (dashed black line) are not the same as the ones presented in \cite{bahar_srgerosita_2024} since the authors there calculate average electron density and temperature profiles which are integrated up to the given radius. In contrast, we calculate the single mass-weighted temperature and electron density for each halo. We account for the redshift evolution including the dependence on the self-similar growth parameter $E(z)=[\Omega_M(1+z)^3 + \Omega_{\Lambda}]^{1/2}$ in the y-axis. The shaded bands represent the $25^{th}-75^{th}$ percentile from the sample. 
\par
The observational data agrees with the predictions from simulations for entropy within $R_{2500}$ and $R_{500}$ within the confidence interval. Furthermore, the difference between the detected and undetected population is negligible for most of the temperature range, or at least within the uncertainty, for these radial ranges. On the other hand, we notice a different trend for the very inner core (i.e., $0.15 R_{500}$) where the undetected and detected populations seem to differ with decreasing temperatures (i.e., smaller mass) even if carrying a large scatter. The detected population is still above the observational data but we notice a much better agreement when we consider the possibility of a selection effect due to the differences in gas fractions at fixed halo mass. Considering that the observational data reported in \cite{bahar_srgerosita_2024} refers to detections with exposure time like eRASS1, we can expect less bright objects (which we can detect in our configuration) to be undetected in the observational analysis, further lowering the plotted profiles. Furthermore, they include a different extent likelihood (i.e., $\mathcal{L}_{EXT}>3$), less conservative than our choice (i.e., $\mathcal{L}_{EXT}>6$). The origins and potential causes for such low gas fractions in the undetected population will be the object of future studies. 

\section{Summary}
\label{sec:conclusions}
eROSITA will give an unprecedented look at the galaxy group mass regime in X-ray observations. It will characterise the thermodynamic properties of the IGrM and provide insightful advancements in our knowledge of the baryonic mass in our Universe. To support such observational effort, a direct and meaningful comparison with the state-of-the-art cosmological simulations is necessary to assess and discuss the systematics present in our investigations. 
\par
The results presented in this study are based on the \textit{Magneticum Pathfinder}, a publicly available set of cosmological hydrodynamical simulations \citep{hirschmann_cosmological_2014}, which is used to construct synthetic X-ray observations in a $30\times 30$ deg$^2$ lightcone down to redshift $z=0.2$. The X-ray photons are extracted thanks to PHOX \citep{biffi_observing_2012, biffi_investigating_2013, biffi_agn_2018, vladutescu-zopp_decomposition_2023} which models the emission from hot gas, AGNs, and XRBs, considering absorption processes and spectral characteristics. Event files are then created with SIXTE \citep{dauser_sixte_2019}, a software package simulating the detection process and background components in a mock eROSITA observation with the first 4 eROSITA all-sky surveys combined (eRASS:4) depth \citep{predehl_erosita_2021}. The modelling of foreground and background emissions, including contributions from unresolved sources and Galactic components, is discussed in detail and adapted from the recent observational results \citep{liu_erosita_2022, ponti_abundance_2023}. We run the halo finder on the event files to reconstruct the extended and point source catalogues in the $0.2-2.3$ keV band. 
We follow the prescriptions adopted for eFEDS \citep{predehl_erosita_2021,liu_2022groupsclusters} to determine the likelihood limits after scanning the effects on completeness and contamination when releasing such constraints. This allows us to derive a matched catalogue of extended sources from which extracting the group selection function in our mock observations. We proceed to do the same for the point sources with the AGN population.
\par
In the following, we summarise our main findings.
\begin{itemize}
    \item As expected, the eSASS detections are composed mostly of point sources (i.e., AGNs, misclassified faint halos, fluctuations in the background) and for 6\% of extended sources. Matching according to the spatial distribution and high flux of the extended sources reveals the completeness of the survey as a function of the input flux and mass as shown in Fig.~\ref{fig:luminosity_z}. This corresponds to a contamination of 3\% in the extended eSASS catalogue. Misclassification of extended-to-point sources is significant at low fluxes (or low mass), as shown in Fig.~\ref{fig:completeness}. Blended sources only sum up to 5\% (or less including a second match with the point source catalogue) thus it does not represent our largest source of misclassification which is fragmentation (12\%) of up-close sources. We find that fragmentation is more prone in high-flux sources ($>10^{-12}$ erg~s$^{-1}$~cm$^{-2}$) rather than extended ones. 
    \item Luminosities of extended detection are extracted through flux aperture estimation within the input halo $R_{500}$ radius. We recover unbiased luminosities, despite a mild scatter which increases (reaching $\sim 20$\%) towards lower luminosities, as illustrated in Fig.~\ref{fig:LumVSLum}. We argue that this regime corresponds to halos with low number counts affecting our spectral reconstruction. Assuming a 50\% completeness to determine the selection function leads us to determine a survey flux limit of $9.03\times 10^{-14}$ erg~s$^{-1}$~cm$^{-2}$. Fig.~\ref{fig:luminosity_z} shows the outcome in terms of the luminosity versus redshift distribution of the clusters and groups of galaxies.
    \item This flux limit is used to derive the non-parametric (and unbiased) XLF in the lightcone and compare it to the eFEDS one, finding excellent agreement among them. Fig.~\ref{fig:dndL_x} presents the final shape of the XLF with some comparisons drawn from the literature in the nearby Universe.  
    \item We discuss the use of scaling relations to recover the halo mass of the extended detections, providing a 1:1 comparison with the input mass $M_{500}$ in Fig.~\ref{fig:MassVSMass}. We show that deriving halo masses from luminosities can yield large uncertainties due to the large scatter in the scaling relation.  
    \item In Fig.~\ref{fig:dndL_x_AGN_eSASS} we present the XLF reconstructed for the detected AGN population compared to the one extracted from the mock catalogue described in \cite{marchesi_mock_2020}. Our results show that the bright end of the XLF ($>10^{43}$ erg~s$^{-1}$) is fully consistent with these results, whereas for lower luminosities we find some tensions. 
    \item We discuss the selection effects in an eRASS:4-like observation derived from the different gas fractions at fixed halo mass which leaves a significant imprint in the X-ray emissivity (see Fig.~\ref{fig:SB}--\ref{fig:entropy} and discussion there). Such an effect leads to more easily detecting halos with low entropy corresponding to less centrally concentrated and gas-richer systems. 
\end{itemize}
Thus, we further confirmed that eROSITA is suitable for detecting clusters of galaxies (i.e., $M>10^{14}M_{\odot}$) with high completeness and purity, especially when supplementing the extended detections with matches from the point source catalogue \citep{bulbul_erosita_2022}, and AGNs. As the halo mass goes down completeness follows, therefore in the group regime, it is advisable to complement X-ray observations characterising the hot gas with other wavelengths (e.g., optical) and avoid samples affected by Malqumist bias. Galaxy group detection suffers from significant selection effects, leading to samples biased towards high X-ray luminous halos. In this regime, we suggest stacking as a viable way to shed light on the bulk properties of groups \citep{popesso_x-ray_2023}. Mapping the distribution of baryonic mass in our Universe down to the group-sized halo masses it is essential to clarify how much of the missing baryons is locked up in halos and filaments \citep[see][]{oppenheimer_simulating_2021} and the impact of the baryonic cycle in structure formation \citep{voit_tracing_2005, eckert_feedback_2021} in a large cosmological context. In this scenario, one key role is played by the SMBH sitting at the halo centre and its co-evolution with the host halo. Nowadays, this also represents one of the major theoretical unknowns in our paradigm of galaxy formation and evolution. Numerical simulations implement different AGN feedback models \citep{voit_precipitation-regulated_2015, voit_global_2017, gaspari_linking_2020} which effectively are most efficient at these scales and can even lead to a depletion of baryons within the halo radius.

\begin{acknowledgements}
      We thank Stefano Marchesi and Roberto Gilli for their help on the AGN cosmic background and Nicola Locatelli for the useful discussion on the Galactic foreground emission. We would also like to show our gratitude to Andrea Biviano for the invaluable comments on the draft. IM and VT acknowledge support from the European Research Council (ERC) under the European Union’s Horizon Europe research and innovation programme ERC CoG (Grant agreement No. 101045437, PI P. Popesso). KD acknowledges support by the COMPLEX project from the ERC under the European Union’s Horizon 2020 research and innovation program grant agreement ERC-2019-AdG 882679. VB and SVZ acknowledge support by the DFG project nr. 415510302. NM acknowledges funding by the European Union through a Marie Sk{\l}odowska-Curie Action Postdoctoral Fellowship (Grant Agreement: 101061448, project: MEMORY). YZ and GP acknowledge financial support from the European Research Council (ERC) under the European Union’s Horizon 2020 research and innovation program HotMilk (grant agreement No. 865637). GP acknowledges support from Bando per il Finanziamento della Ricerca Fondamentale 2022 dell’Istituto Nazionale di Astrofisica (INAF): GO Large program and from the Framework per l’Attrazione e il Rafforzamento delle Eccellenze (FARE) per la ricerca in Italia (R20L5S39T9). The calculations for the {\it Magneticum} simulations were carried out at the Leibniz Supercomputer Center (LRZ) under project pr83li. 
\end{acknowledgements}

%
   \bibliographystyle{aa} 
   \bibliography{references} 
%
\begin{appendix}
\section{On the impact of extension in aperture fluxes}
\label{sec:appendix}
\begin{figure*}
    \centering
    \includegraphics[scale=0.5]{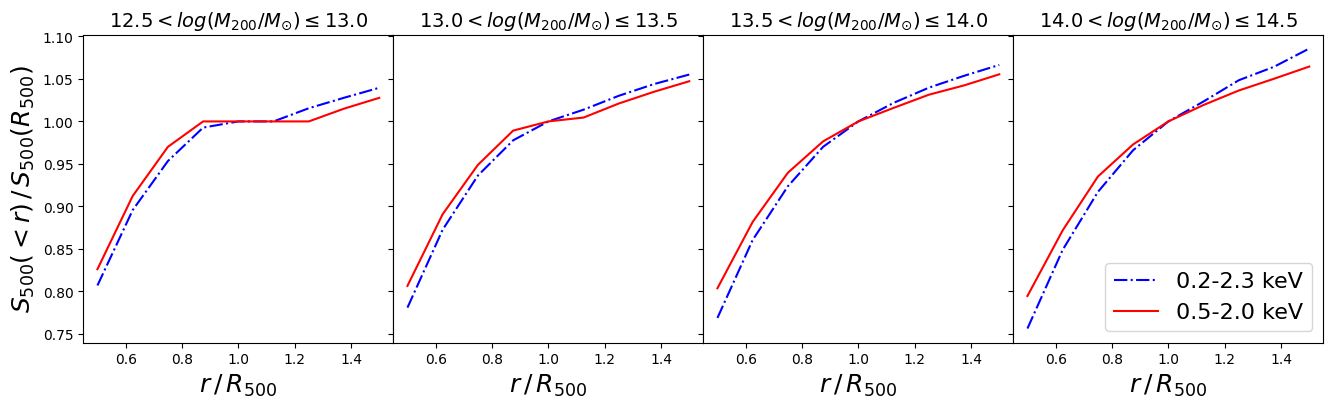}
    \caption{Fraction of the flux within different apertures over the flux in $R_{500}$. The estimates are performed for two energy bands: $0.5-2.0$ keV plotted with a red solid line and $0.2-2.3$ keV with a dash-dotted blue line. The plots represent the medians obtained within the different halo mass bins. }
    \label{fig:flux_in_r500}
\end{figure*}

In this section, we briefly discuss the predictions from our cosmological simulation on the flux estimates from aperture photometry when the assumed radius is incorrectly identified. Fig.~\ref{fig:flux_in_r500} shows the fraction of cumulative X-ray aperture flux in two different soft bands (i.e., $0.5-2.0$ keV and $0.2-2.3$ keV) normalised for the flux within $R_{500}$. The lines are the median drawn from four halo mass bins to properly highlight any possible dependence on the halo size. In the low mass bins, we observe a plateau between $0.8-1.2.0$ $R_{500}$ which suggests that the effects in this regime would not be large, especially for the energy band $0.5-2.0$ keV. At the highest mass bins, recovering the wrong radius can lead up to $\pm 5\% $ difference in the flux estimation within $0.8-1.2 \, R_{500}$, a margin that is small but should still be taken into account when propagating the errors. Extending outside of these ranges can lead to far worse underestimation. Although we do not report it in the figure, we test the energy range $0.1-2.4$ keV which overlaps with the $0.2-2.3$ keV results.  These estimates will depend on the surface brightness profiles of these systems, and thus shallow profiles are expected to be less problematic. 
\end{appendix}

\end{document}